\documentstyle[12pt]{article}
\renewcommand{\thefootnote}{\fnsymbol{footnote}}
\newcommand{\newsection}{
\setcounter{equation}{0}
\section}

\def\appendix#1{
  \addtocounter{section}{1}
  \setcounter{equation}{0}
  \renewcommand{\thesection}{\Alph{section}}
 \section*{Appendix \thesection\protect\indent \parbox[t]{11.715cm} {#1}}
  \addcontentsline{toc}{section}{Appendix \thesection\ \ \ #1}
  }

\newcommand{\tr}[1]{\:{\rm tr}\,#1}
\def\e{{\, e}\,}
\newcommand{\rf}[1]{(\ref{#1})}
\newcommand{\eq}[1]{eq.~(\ref{#1})}
\newcommand{\non}{\nonumber \\*}
\def\be{\begin{equation}}
\def\ee{\end{equation}}
\def\bea{\begin{eqnarray}}
\def\eea{\end{eqnarray}}
\def\seff{S_{\rm eff}}
\def\tw{\tilde{W}}
\def\gii{\sqrt{\vphantom{g_j}\alpha \beta g_i+F^2}}
\def\gjj{\sqrt{\alpha \beta g_j+F^2}}
\def\fii{\sqrt{\vphantom{\lambda_j}a\lambda_i+\eta ^2}}
\def\fjj{\sqrt{a\lambda _j+\eta ^2}}
\def\const{{\rm const}}
\def\fdet{\mathop{\rm Det}}

\begin{document}
\begin{titlepage}
\begin{flushright}
ITEP--TH--18/97\\
hep-th/9705014\\
\end{flushright}
\vspace{.5cm}

\begin{center}
{\LARGE Effective Action and Measure  \\[.4cm]
 in Matrix Model of IIB Superstrings }\\
\vspace{1.2cm}
{\large L. Chekhov${}^{{\rm a)}}$\footnote{E-mail: chekhov@genesis.mi.ras.ru}
 and K. Zarembo${}^{{\rm b)}}$%
\footnote{E-mail: zarembo@vxitep.itep.ru} }\\
\vspace{24pt}
${}^{{\rm a)}}${\it Steklov Mathematical Institute,}
\\ {\it  Gubkina 8, 117966, GSP--1, Moscow, Russia} \\[.2cm]
${}^{{\rm b)}}${\it Institute of Theoretical and Experimental Physics,}
\\ {\it B. Cheremushkinskaya 25, 117259 Moscow, Russia}
\end{center}
\vskip 0.9 cm
\begin{abstract}
 We calculate an effective action and measure induced by the
 integration over the auxiliary field in the matrix model recently
 proposed to describe IIB superstrings. It is shown that the measure of
 integration over the auxiliary matrix is uniquely determined by locality
 and reparametrization invariance of the resulting effective action.
 The large--$N$ limit of the induced measure for string coordinates 
 is discussed in detail. It is found to be ultralocal and, thus, 
 possibly is irrelevant in the continuum limit.
 The model of the GKM type is considered in relation
to the effective action problem.
\end{abstract}
\end{titlepage}
\setcounter{page}{1}
\renewcommand{\thefootnote}{\arabic{footnote}}
\setcounter{footnote}{0}

\newsection{Introduction}

  The matrix model formulation of M theory \cite{BFSS96} renewed an
  interest in the problem of a nonperturbative approach to
  superstrings.   M(atrix) model \cite{BFSS96} is expected
  to provide a complete quantum-mechanical description of M theory
  thus unifying all known string theories. The emergence of
  perturbative strings in M(atrix) theory is discussed in a
  series of recent papers \cite{strings}.

  In  the other approach \cite{IKKT}, the matrix model is directly
  identified with type IIB superstring.
  The action of this model is obtained from the
  ten-dimensional $U(N)$ super Yang--Mills theory by the reduction
  to a point. The
 string action follows from the matrix model by substituting
 the Poisson brackets for the commutators and the
 integrals  over the world sheet for the traces.
This is justified at infinite
 $N$ if the large $N$ limit is identified with the semiclassical
 one. As a result of this procedure, the Schild
  action of IIB Green--Schwarz superstring with fixed
  $\kappa$--symmetry arises:
  \be S_{\rm Schild} =\int d^2\sigma \left(
   \frac{\alpha }{4\sqrt{g}}\{X^{\mu},X^{\nu}\}_{PB}^2
-\frac{i}{2}\bar{\psi}\Gamma^{\mu}\{X_{\mu},\psi\}_{PB}
+\beta \sqrt{g}\right),
\label{Saction}
\ee
 where $\{\,,\,\}_{PB}$ are the usual Poisson brackets.

 The Schild action \rf{Saction} is equivalent to the Nambu--Goto action,
 if the auxiliary field $\sqrt{g}$ is excluded via its
 classical equations of motion.
 On the quantum level, one should
 average w.r.t.\ $\sqrt{g}$ in the functional integral.  The
 arguments have been given that the Polyakov string is reproduced in
 this manner, provided that the measure of integration is chosen
 properly and conformal anomaly cancels \cite{Yoneya}.

 In the IKKT approach
 \cite{IKKT}, the integration over $\sqrt{g}$ is modelled by a summation
 over the matrix size $N$, which, therefore, is a dynamical variable.
 A modification of the IKKT model was proposed \cite{FMOSZ}
 with an additional matrix variable $Y$ introduced,
 which enters the action of the matrix model the same way as $\sqrt{g}$
 enters the string action:
\begin{equation}
S_{NBI}=-\frac{\alpha}{4} \tr Y^{-1}[A_\mu,A_\nu]^2+\beta \tr Y
 -\frac{1}{2} \tr \bar{\psi}\Gamma^\mu[A_\mu,\psi].
\label{p1}
\end{equation}
 There is no summation over $N$ in this model and the large $N$ limit
 can be taken straightforwardly, yielding the string action \rf{Saction}
 in the semiclassical (large $N$) approximation.

 The main idea of the matrix model approach is to consider the matrix
 integral in the infinite $N$ limit as the nonperturbative definition
 of the string partition function. The crucial point is the
 choice of an integration measure, which should respect all the
 symmetries of the underlying string theory. In \cite{FMOSZ}, it was
 suggested to define the partition function as follows:
 \begin{equation}\label{nbi}
 Z_{NBI}=\int dA_\mu\, d\bar{\psi }\,d\psi
 \,dY\,(\det Y)^{-\gamma }\,\, \e^{-S_{NBI}}.
 \end{equation}
 The integration over $Y$  ranges over positive Hermitean matrices.
 In this paper, we focus on the result of this integration, which
 yields an
 effective action and measure for $A_\mu $ and their superpartners:
 \begin{equation}\label{effact}
 J(G)\e^{-\seff(G)
 +\frac{1}{2} \tr \bar{\psi}\Gamma^\mu[A_\mu,\psi]}
 =\int dY\,(\det Y)^{-\gamma }\,\,\e^{-S_{NBI}},
 \end{equation}
 where
 \begin{equation}\label{defg}
 G=-[A_\mu ,A_\nu ]^2
 \end{equation}
 is a Hermitean positively definite matrix. At large $N$, when the
 commutator is replaced by the Poisson bracket, $G$ becomes the
 determinant of an induced metric on the world sheet:
 \begin{equation}\label{gninf}
 G\longrightarrow \{X_\mu ,X_\nu \}_{PB}^2=
 2\det\limits_{ab}\partial _aX_\mu
 \partial _bX^\mu.
 \end{equation}
 Note that $Y$ does not
 interact with fermions and the corresponding term in the action remains
 unchanged.
 
 The separation of the result of integration over an auxiliuaty field $Y$ 
 into the effective action and measure is to some extent arbitrary. As a rule,
 all power-like terms in the exponential are naturally interpreted as an 
 effective action. We also encounter the logarithmic terms which we refer
 to the measure of integration over $A_\mu$. Such interpretation
 is justified in the large--$N$ limit, as discussed in sec.~4.

 It is evident that in the saddle point approximation the effective
 action does not differ from the matrix counterpart of the Nambu--Goto one:
 \begin{equation}\label{ng}
 S_{NG}=\sqrt{\alpha \beta }\tr\sqrt{G}=\sqrt{\alpha \beta
 }\tr\sqrt{-[A_\mu ,A_\nu ]^2},
 \end{equation}
 If $A_\mu $ are regarded
 as a zero dimensional gauge potentials, this expression looks like the
 strong coupling non-Abelian Born--Infeld (NBI) action.
 The Nambu--Goto action is reproduced from the NBI
 matrix model \rf{p1}, \rf{nbi} even quantum--mechanically, up to a
 measure factor, for the special value of $\gamma =N-1/2$, where the
 integration over $Y$ can be performed explicitly \cite{FMOSZ}.  Our main
 goal is to check by the direct calculation, whether or not this result
 is valid in general, and how does the effective action look for various
 values of $\gamma $.

 We demonstrate that $\gamma =N+O(1)$ is singled out
 by general principles of locality and reparametrization
 invariance of the string action.
 It seems, however,
 interesting to investigate the model (\ref{nbi}) at arbitrary $\gamma$,
 since this may correspond to some soft breaking of the reparametrization
 invariance in the theory.
 It is worth mentioning that the value of $\gamma=N+O(1) $ is
 distinguished from many points of view. It was found to be critical for
 the matrix model, obtained from \rf{nbi} by excluding $A_\mu $ via their
 classical equations of motion \cite{KO}. On the other hand,
 the integral over the Hermitean matrix $Y$ with $A_\mu $ treated as the
 external fields was shown \cite{MMS} to be equivalent for $\gamma =N$ to
 the unitary matrix integral of \cite{unit,BG}.

 In the
 general case, the integral over $Y$ can not be calculated exactly and we
 use the methods of the large $N$ expansion, systematically dropping the
 corrections in $1/N$.  The parameters $\gamma $, $\alpha $ and $\beta
 $ are assumed to be of order $N$.

\newsection{Matrix model}

 According to the definition of the effective action~\rf{effact},
 \begin{equation}\label{matmod}
 Z(G)\equiv J(G)\e^{-\seff(G)}=\int dY\,\e^{-\frac{\alpha }{4}\tr
 Y^{-1}G-\beta \tr Y-\gamma \tr \log Y}.
 \end{equation}
 This integral can be viewed as a one--matrix model with the external
 field $G$. It is convenient to make the change of
 the integration variables: $Y=\frac{N}{\beta }X^{-1}$, then
 $dY\propto(\det X)^{-2N}dX$ and, up to an irrelevant constant,
 \eq{matmod} can be rewritten in the following form:
 \begin{equation}\label{mm}
 Z=\int dX\,\e^{-N\tr\left[X\Lambda +X^{-1}+(2\eta +1)\log X\right]},
 \end{equation}
 where we have introduced the notations:
 \begin{equation}\label{not1}
 \Lambda =\frac{\alpha \beta }{4N^2}\,G
 \end{equation}
 and
 \begin{equation}\label{not2}
 \eta =\frac{1}{2}\left(1-\frac{\gamma }{N}\right).
 \end{equation}
 The constant $\eta $ is normalized to be of order unity and to
 vanish for the critical value of $\gamma$.

 The matrix integral \rf{mm} belongs to a class of generalized Kontsevich
 models \cite{GKM}. Such models with negative powers of the matrix~$X$
 have been previously discussed in the context of
 $c=1$ bosonic string theory \cite{DMP}. In \cite{MMS}, the $\tau$-function
 approach to such models was developed. There, the parameter~$\eta$ plays
 the role of the zeroth time in the corresponding integrable hierarchy.
 Moreover, at the conformal point $\eta=0$, this model was shown
 \cite{MMS} to have the same Schwinger--Dyson equations as the $U(N)$
 model solved in \cite{unit,BG}.

 For the models of this type, the
 large $N$ solution is known explicitly only
 in some special cases.  The models with cubic potential for $X$
 \cite{cubic} and the combination of the logarithmic and quadratic
 potentials \cite{CM,ZC} were solved by a method based on Schwinger--Dyson
 equations, developed first for the unitary matrix models with external
 field \cite{unit,BG}.
 The same technique, being applied to the
 integral \rf{mm}, also allows to find its large $N$ asymptotics in
 the closed form for arbitrary $\eta $.

  The Schwinger--Dyson equations for \rf{mm} follow from the identity
 \begin{equation}\label{...=0}
 \frac{1}{N^3}\,\frac{\partial }{\partial \Lambda_{jk} }
 \,\frac{\partial }{\partial \Lambda_{li} }\,\int dX\,\,
 \frac{\partial }{\partial X_{ij}}\,
 \e^{-N\tr\left[X\Lambda +X^{-1}+(2\eta +1)\log X\right]}=0 .
 \end{equation}
 This identity gives rise to the differential equation for
 $Z(\Lambda )$:
 \begin{equation}\label{diffSD}
 \left[-\frac{1}{N^2}\,\Lambda _{ji}\,\frac{\partial
 }{\partial \Lambda _{jk}}\,\frac{\partial }{\partial
 \Lambda _{li}}
 +\frac{1}{N}\,(2\eta -1)\,\frac{\partial }{\partial
 \Lambda _{lk}}
 +\delta _{kl}\right]Z(\Lambda )=0.
 \end{equation}
 Due to the invariance of the integration measure and the
 action under the unitary transformations, the partition function $Z$
 depends only on the eigenvalues $\lambda _i$ of the matrix $\Lambda $
 and is symmetric under their permutations. Hence, only $N$ of the $N^2$
 Schwinger--Dyson equations \rf{diffSD} are linearly independent. Being
 written in terms of the eigenvalues, these $N$ equations read
 \begin{equation}\label{SDeig}
 \left[-\frac{1}{N^2}\,\lambda _i\,\frac{\partial ^2}{\partial
 \lambda_i^2} -\frac{1}{N^2}\,\sum_{j\neq i}\lambda _j\,\frac{1}{\lambda
 _j-\lambda _i}\,\left(\frac{\partial}{\partial \lambda _j}-
 \frac{\partial }{\partial \lambda _i}\right)
 +\frac{1}{N}\,(2\eta -1)\,\frac{\partial
 }{\partial \lambda _i}+1\right]Z(\lambda )=0.
 \end{equation}
 For $\eta=0$, these formulas coincide with the
 corresponding formulas for the $U(N)$ model~\cite{unit,BG}.
 Here, we solve this model for the case $\eta\ne0$.

 It is convenient to set
 \begin{equation}\label{defW}
 W(\lambda _i)=\frac{1}{N}\,\frac{\partial }{\partial \lambda _i}\,\log Z.
 \end{equation}
 We also introduce the eigenvalue density of the matrix $\Lambda $:
 \begin{equation}\label{dens}
 \rho (x)=\frac{1}{N}\,\sum_{i}\delta (x-\lambda _i).
 \end{equation}
 The density obeys the normalization condition
 \begin{equation}\label{norm}
 \int dx\,\rho (x)=1
 \end{equation}
 and in the large $N$ limit becomes a smooth function.

 A simple power counting shows that the derivative of $W(\lambda_i )$ in
 the first term on the left hand side of equation \rf{SDeig} is suppressed
 by the factor $1/N$ and can be omitted at $N=\infty $. The remaining
 terms are rewritten as follows:
 \begin{equation}\label{inteqn}
 -xW^2(x)-\int dy\rho (y)\,y\,\frac{W(y)-W(x)}{y-x}+(2\eta -1)W(x)+1=0,
 \end{equation}
 where $\lambda _i$ is replaced by $x$. The equation \rf{inteqn} can be
 simplified by the substitution
 \begin{equation}\label{subst}
 \tw(x)=xW(x)-\eta.
 \end{equation}
 After some transformations, using the normalization condition \rf{norm},
 we obtain
 \begin{equation}\label{maineq}
 \tw^2(x)+x\int dy\rho (y)\,\,\frac{\tw(y)-\tw(x)}{y-x}=x+\eta ^2.
 \end{equation}

 The nonlinear integral equation \rf{maineq} can be solved with the
 help of the anzatz
 \begin{equation}\label{anz}
 \tw(x)=f(x)+\frac{x}{2}\,\int dy\,\frac{\rho
 (y)}{f(y)}\,\frac{f(y)-f(x)}{y-x},
 \end{equation}
 where $f(x)$ is an unknown function to be determined by substituting
 \rf{anz} into \eq{maineq}. The asymptotic behaviors of $\tw(x)$ and
 $f(x)$ as $x\rightarrow \infty $ follow from eq.~\rf{maineq}:
 $\tw(x)\sim \sqrt{x}+1/2$, and the analytic solution with
 minimal set of singularities is simply
 \begin{equation}\label{fx}
 f(x)=\sqrt{ax+b}.
 \end{equation}
 The parameters $a$ and $b$ are unambiguously determined from
 \eq{maineq}. We find that $b=\eta ^2$ and $a$ is implicitly  defined by
 the equality
 \begin{equation}\label{defa}
 1+\frac{1}{2}\int dy\,\frac{\rho (y)}{f(y)}=\frac{1}{\sqrt{a}},
 \end{equation}
 or, in terms of the eigenvalues,
 \begin{equation}\label{a1}
 1+\frac{1}{2N}\sum_{j}\frac{1}{\sqrt{a\lambda _j+\eta
 ^2}}=\frac{1}{\sqrt{a}}.
 \end{equation}

 For the logarithmic derivative of the partition function, $W(\lambda
 _i)$, we have, according to eqs.~\rf{subst} and \rf{anz}:
 \begin{equation}\label{wx}
 W(\lambda _i)=\frac{\eta }{\lambda _i}+\frac{\sqrt{a\lambda _i+\eta
 ^2}}{\lambda _i} +\frac{a}{2N}\sum_{j}\frac{1}{\sqrt{a\lambda _j+\eta
 ^2}}\, \frac{1}{\sqrt{a\lambda _j+\eta
 ^2}+\sqrt{\vphantom{\lambda _j}a\lambda _i+\eta ^2}}.
 \end{equation}
 To calculate the partition function $Z(\lambda )$ one should
 integrate eq.~\rf{defW}.  This
 integration is complicated by the fact that $a$ is also the function of
 $\lambda _i$.  However, the problem can be avoided by the following
 trick. We integrate eq.~\rf{wx} w.r.t.\ $\lambda _i$ as if $a$
 is a constant. Then we can add an
 arbitrary function of $a$ to the expression obtained.
 The proper choice of this function makes the
 final expression stationary in $a$. Thus, we obtain:
 \begin{eqnarray}\label{res}
 \log Z&=&N^2\left[\left(\eta ^2+\frac{1}{4}\right)\log a+\frac{4\eta
 ^2}{\sqrt{a}}-\frac{\eta ^2}{a}\right]
 \non  &&
 +N\sum_{i}\left[
 \frac{2}{\sqrt{a}}\,\fii+\eta \log\left(\lambda _i\,\frac{\fii-\eta
 }{\fii+\eta }\right)\right]
 \non  &&
 -\frac{1}{2}\sum_{ij}\log\left(\fii+\fjj\right).
 \end{eqnarray}
 One can verify directly
 that $\frac{\partial }{\partial
 a}\log Z=0$ and $\frac{1}{N}\,\frac{\partial }{\partial \lambda _i}\log
 Z=W(\lambda _i)$, as far as \eq{a1} holds.

\newsection{Effective action and measure}

 We refer all logarithmic terms in \rf{res} to the induced measure. 
 Then the remaining terms represent the effective action.
 Returning to the original notations \rf{not1}, \rf{not2},  after
 some transformations we find from \eq{res}:
 \begin{equation}\label{se}
 \seff(G)=\tr\sqrt{\alpha \beta G+F^2}+(N-\gamma )F-\frac{1}{4}\,F^2,
 \end{equation}
 \begin{equation}\label{me}
 J(G)=F^{-\frac{1}{2}(N-\gamma
 )^2}\prod_{i}\left(g_i\,\frac{\gii-F}{\gii+F}\right)^{\frac{1}{2}(N-\gamma )}
 \prod_{i<j}\frac{1}{\gii+\gjj}.
 \end{equation}
 Here $g_i$ are the eigenvalues of the matrix $G$. The parameter $F$ is
 related to $a$ by
 \begin{equation}\label{deff}
 F^2=\frac{4\eta ^2N^2}{a}.
 \end{equation}
 It can be found from \eq{a1}, which is rewritten as follows:
 \begin{equation}\label{f1}
 F=N-\gamma +F\tr\,\frac{1}{\sqrt{\alpha \beta G+F^2}}\,.
 \end{equation}

 General expressions \rf{se} and \rf{me} considerably simplify
 for $\gamma =N+O(1)$. In this case, $F\rightarrow 0$ and \rf{se} reduces
 to the Nambu--Goto action \rf{ng}, as expected from \cite{FMOSZ}. We
 also find that this value of $\gamma $ is critical for the matrix
 model involved -- the measure  $J(G)$ is nonanalytic at $\gamma =N$:
 \begin{equation}\label{crit}
 J(G)=\e^{-\frac{1}{2}(N-\gamma )^2\log(N-\gamma) }\times{\rm regular}.
 \end{equation}
 The critical behavior we have found is typical for all matrix models
 with logarithmic potentials \cite{log,CM} and, probably, it is the same
 that was discussed in \cite{KO}. This nonanalyticity of the measure
 \rf{crit} is harmless -- the limit at $\gamma=N$ exists. Moreover, this
 singularity seems to be irrelevant for the description of superstrings,
 since the nonanalytic piece of the measure factorizes out and does not
 depend on $G$, hence, it cancels in all correlation functions of $A_\mu $
 and $\psi $.

 At the critical point, we have, dropping an overall constant factor,
 \begin{equation}\label{mc}
 J(G)=\prod_{i<j}\frac{1}{\sqrt{\vphantom{g_j}g_i}+\sqrt{g_j}}.
 \end{equation}
 In fact, for $N=\gamma -1/2$ this result is exact, i.e., it is valid
  beyond the large $N$ approximation \cite{FMOSZ}. It is interesting to
  compare \rf{mc} and \rf{se} at $F=0$ with the solution of the unitary
  matrix model \cite{unit,BG}. In accord with \cite{MMS}, we find the
  agreement with the $U(N)$ model in the weak coupling
  (large $G$) phase \cite{unit,BG}. The strong coupling (small $G$) phase
  \cite{BG} of the unitary model is characterized by the partition function
  $Z(G)$, which is analytic at $G=0$, and this phase
  corresponds to the different branch of the solution to
  \eq{f1}. However, it is never realized for the Hermitean model,
  since the matrix $G$ is responsible for the convergence of the
  integral, and one can not expand it in $G$, whatever small it is. In
  the case of the unitary matrix model, there are no problems with
  convergence because of the compactness of the integration domain.

\newsection{Discussion}

 In the large $N$ limit the matrix $G$ is replaced by the determinant of
 the induced metric on the string world sheet according to
 eq.~\rf{gninf}.  Let us consider the induced measure \rf{mc} in this
 limit\footnote{We are thankful to P.~Olesen for the discussion of this
 point.}.  For this purpose, we rewrite \eq{mc} in the form
 \begin{equation}\label{mc'} J(G)=\left[\frac{\Delta^2
 (\sqrt{g})}{\Delta^2 (g)}\right]^{1/2}=\left(\frac{\det'
 \left[\sqrt{G},\cdot\,\right]}{\det' [G,\cdot\,]}\right)^{1/2}\,,
 \end{equation}
 where $\Delta $ is the Vandermonde determinant,
 $\Delta (g)=\prod\limits_{i<j}(g_i-g_j)$, and the prime denotes that
 zero modes are omitted in the determinants.  Replacing the commutators by
 the Poisson brackets, we obtain:
 \begin{equation}\label{largeN}
 J[G]=\left(\frac{\fdet'
 \left\{\sqrt{G},\cdot\,\right\}_{PB}}{\fdet'\{G,\cdot\,\}_{PB}}
 \right)^{1/2}=\left(\frac{\fdet'\varepsilon ^{ab}\partial
 _a\sqrt{G}\partial _b} {\fdet'\varepsilon ^{ab}\partial _aG\partial
 _b}\right)^{1/2}\,.
 \end{equation}
 This fraction of the determinants can be substantially simplified, since
 \begin{equation}
 \varepsilon ^{ab}\partial
 _a\sqrt{G}\partial _b=\frac{1}{2\sqrt{G}}
 \varepsilon ^{ab}\partial _aG\partial_b.
 \end{equation}
 Therefore,
 \begin{equation}\label{mn}
 J[G]=\const\,\left(\fdet \sqrt{G}\,\right)^{-1/2}.
 \end{equation}
 This means that the measure of integration over the string coordinates
 has the form:
 \begin{equation}
 \label{DX}
 DX_{\mu}=\prod_{\sigma}dX_\mu(\sigma)G^{-1/4}(\sigma).
 \end{equation}
 This expression confirms our interpretation of $J(G)$ as the induced
 measure factor, rather than the part of the effective action.

 Some comments concerning the measure \rf{DX}
 are in order. First of all, the
 measure \rf{DX} respects the reparametrization invariance, because
 under the world sheet diffeomorphisms $\sigma \rightarrow \sigma '$
 the measure is multiplied by the (infinite) constant:
 \begin{equation}
 DX_{\mu}\rightarrow\const\, DX_{\mu}.
 \end{equation}
 Since this constant does not depend on the fields, it cancels in all
 correlation functions. Another important property of the induced measure
 is its ultralocality:
\begin{equation}
J[G]=\exp\left(-\frac{1}{4}
\int d^2\sigma\,\delta (\sigma -\sigma ')\biggr|_{\sigma
'\rightarrow\sigma } \ln G(\sigma)\right).
\end{equation}
 In some regularizations $\delta (0)=0$. So, probably the extra factor in
 the measure can be even omitted. It is worth mentioning that
 the replacement of the string coordinates by the matrices already 
 introduces
 some kind of momentum cutoff. The number of the Fourier harmonics in 
 the matrix regularization is
 effectively of order $N$, consequently $N$ plays the role of the largest
 possible momentum. Thus, in the matrix regularization $\delta (0)=N^2$.

 It seems likely that the matrix model \rf{matmod} deserves the
 investigation. Its partition function can be developed into the power
 series in $1/N$ similar to the solutions obtained in \cite{IZ92}
 for the Kontsevich matrix model and in \cite{ACKM} for the
 Kontsevich--Penner (Hermitian one-matrix) model. It would be interesting
 to find whether the model \rf{matmod} has a geometrical meaning from
 the point of view of the moduli spaces (as the two previous models have).

 However, from the string theory point of view,
 the effective action \rf{se} looks rather complicated unless $\gamma
 =N+O(1)$.  After replacing the traces and the commutators by the
 integrals and the Poisson brackets it should be interpreted as a world
 sheet action of the string.  In general, this action  is nonlocal due to
 the implicit dependence of the parameter $F$ on $G$.  What is more
 important, it is not reparamerization invariant -- while $\sqrt{G}$
 transforms under general diffeomorphisms as a scalar density, $F$ has no
 definite transformation law, as it can be seen from its defining
 equation \rf{f1}. The only way to bypass these difficulties is to set
 $F=0$, which is only possible for $\gamma =N+O(1)$. Thus, the general
 principles of the string theory select the unique, up to the terms
 subleading in $1/N$, value of $\gamma $.

 Qualitatively, these results can be explained, if we consider the
 scaling transformations: \mbox{$\sigma \rightarrow \lambda \sigma $},
 \mbox{$\{\,,\,\}_{PB}\rightarrow \lambda ^{-2}\{\,,\,\}_{PB}$},
 \mbox{$\sqrt{g}\rightarrow \lambda ^{-2}\sqrt{g}$}.  Strictly speaking,
 they have no analogues in the
 matrix model, since the measure of integration over the world sheet of
 the string, $d^2\sigma $, also should transform. In the matrix model
 this would correspond to the transformation of traces:  $\tr\rightarrow
 \lambda^2 \tr$. However, if the functional measure in the matrix model is
 invariant under the rescaling of $Y$, which is the
 counterpart of $\sqrt{g}$, the value of $\gamma $ is immediately fixed
 to be equal to $N$.

\subsection*{Acknowledgments}

 The authors are grateful to Y.~Makeenko, A.~Mironov and P.~Olesen for
 discussions.  This work was supported in part by RFFI grant
 96--01--00344.  The work of K.Z. was supported in part by CRDF grant
 96--RP1--253, INTAS grant 94--0840 and grant 96--15--96455 of the
 support of scientific schools.

\end{document}